\documentclass[a4paper]{article}
\usepackage{amsfonts}
\usepackage{amssymb}
\usepackage{amsmath}
\usepackage{graphics}
\usepackage{color}
\usepackage{graphicx}

\usepackage{latexsym}
\usepackage{geometry}
\usepackage{verbatim}
\usepackage{authblk}

\newcommand{\bean}{\begin{eqnarray}}
\newcommand{\eean}{\end{eqnarray}}
\newcommand{\eqs}[1]{Eqs. (\ref{#1})}
\newcommand{\eq}[1]{Eq. (\ref{#1})}
\newcommand{\meq}[1]{(\ref{#1})}
\newcommand{\fig}[1]{Fig. \ref{#1}}
\newcommand{\ppa}[2]{\left(\frac{\partial}{\partial #1}\right)^{#2}}
\newcommand{\pp}[2]{\frac{\partial #1}{\partial #2}}

\newcommand{\hsp}{\hspace{0.1mm}}    
\newcommand{\ep}{\epsilon}

\newcommand{\bea}{\begin{eqnarray*}}
\newcommand{\eea}{\end{eqnarray*}}
\newcommand{\grad}{\nabla}

\newcommand{\se}[1]{Section \ref{#1}}

\newcommand{\eqn}{&=&}
\newcommand{\non}{\nonumber \\}
\newcommand{\lb}{\left(}
\newcommand{\rb}{\right)}
\newcommand{\lsb}{\left[}
\newcommand{\rsb}{\right]}
\newcommand{\md}{\mathrm{d}}

\begin{document}
\title{Geodesic completeness, curvature singularities and infinite tidal forces}
\author{Xiaotian Zhang}
\author{Sijie Gao\thanks{Corresponding author: sijie@bnu.edu.cn}}
\affil[1]{School of Physics and Astronomy, Beijing Normal University, Beijing , China,100875}
\date{}

\maketitle

\newtheorem{theo}{Theorem}

\begin{abstract}
We report some new findings regarding the subtle relations among geodesic completeness, curvature singularities and tidal forces. It is well known that any particles may encounter infinite tidal forces near a black hole singularity. However, we find that singularity is not the only cause of  tidal force divergence. Even on the surface of the Earth, the tidal force experienced by a particle could be arbitrarily large if the particle moves arbitrarily close to the speed of light in a nonradial direction. For fixed particle energy, the maximum tidal acceleration occurs for motions parallel to the surface with separation vectors oriented radially.
Recent discoveries of spacetimes in which the metric remains well-defined at curvature singularities have suggested that geodesics might extend through such points. By taking into account the fact that any real particle is an extended body, we calculate the tidal force acting on the particle in a static and spherically symmetric spacetime. We explicitly show that an infinite tidal force always occurs near such a singularity. Therefore, no particle can actually reach the curvature singularity even if the metric is well defined at that point. We also demonstrate that the tidal acceleration along a null geodesic at the coordinate origin is divergent. Finally, we examine a wormhole solution which possesses a curvature singularity at its throat and was previously asserted to be geodesically complete in the literature. However, we prove that no metric can be defined at the throat and thus the spacetime is geodesically incomplete.

\end{abstract}

\section{Introduction}
Singularity is a fundamental prediction of general relativity, with their existence rigorously established by classical singularity theorems \cite{p1965a}-\cite{hp1970}. On the other hand, singularities inevitably lead to pathological behaviors of spacetimes \cite{waldgr}. Efforts to resolve singularity problems have been made by circumventing some of the assumptions in  the singularity theorems\cite{Senovilla:2014gza} or exploring quantum effects to avoid singularities \cite{Hofmann:2015xga}\cite{Casals:2016odj}.

In fact, there are different ways to define singularities\cite{waldgr}. The most common ones include the curvature singularity and geodesic incompleteness. Curvature singularity usually means that the curvature invariants are divergent somewhere. Among all the invariants, seventeen curvature invariants, called Zakhary-Mcintosh (ZM) invariants, form a complete set\cite{Zakhary:1997xas}-\cite{Overduin:2020aiq}. For spherically symmetric spacetimes, it has been shown \cite{Hu:2023iuw}  that the Kretschmann scalar $K=R_{abcd}R^{abcd}$ and the Ricci square $R_{ab}R^{ab}$ are sufficient to determine whether all the seventeen ZM invariants are finite or infinite. Geodesic incompleteness refer to that the affine parameter along a timelike or null geodesic can not be extended to  arbitrarily large values in either the future or past direction.

The two definitions are consistent in many cases, such as the central singularities in the Schwarzschild black hole and in the Kerr black hole. However, counterexamples have been found. For the well known C-metric\cite{weyl}-\cite{ash81}, there exists a conic singularity while all polynomial curvature scalars are finite. On the other hand, spacetimes with curvature singularities are found to be geodesically complete \cite{regume}. This may seem incredible since metric usually cannot exist where a curvature singularity appears. In fact, a regular metric can generate a spacetime with curvature singularities  and a series of solutions have been explicitly  constructed \cite{regume}. Unlike the usual curvature singularity where the metric cannot be defined, the metric is well defined where the curvature blows up. In such spacetimes, particles or light can reach or pass through the singularity. These spacetimes should be distinguished from regular black holes where the metric is regular everywhere and there are no central singularities.  \cite{bardeen}-\cite{miaoreview}.

Despite the controversial definitions of singularities, it is of interest to explore their physical implications. If a particle passes through a curvature singularity, it is natural to ask whether  observational effects would arise. Any real particle is an extended body, rather than a point. It is well known that extended bodies in gravitational field will experience  tidal forces . The tidal force is  an effect of spacetime curvature and plays an important role in astrophysics. It is closely related to gravitational waves\cite{Goswami:2019fyk}.
Meanwhile, a star may be disrupted due to tidal forces produced by a black hole\cite{Kesden:2011ee}, called the tidal disruption event (TDE). TDEs may power bright flares as ultraviolet\cite{Gezari:2006fe} and optical\cite{McKee:2015hwa} radiation. In addition, analyses suggest that tidal accelerations can describe relativistic flow formation\cite{Bini:2017uax}-\cite{Chicone:2003yv}. In previous works, tidal effects have been studied for numerous exact solutions to Einstein's equations, including the Schwarzschild spacetime\cite{Pirani:1956tn}-\cite{Vandeev:2022gbi}, the Reissner-Nordstr\"om  spacetime\cite{Crispino:2016pnv}\cite{Gad:2010ion}, the Kerr spacetime\cite{LimaJunior:2020fhs}-\cite{Chicone:2006rm}, and regular black holes \cite{Sharif:2018gaj}\cite{Lima:2020wcb}.

In this paper, we reveal that the relationships among regular metrics, curvature singularities, and tidal accelerations are far more complex and subtle.
It is commonly believed that infinite curvature would result in infinite tidal forces and vice versa.  In \se{sec-earth}, we explicitly demonstrate that tidal acceleration can become infinite even on the Earth's surface, provided the particle approaches the speed of light very closely. This is essentially the ``Lorentz boost '' effect in gravitational field, which was previously studied in \cite{mat74} \cite{sexl71}, though the tidal effects were only implicitly discussed. We demonstrate that an extended particle moving tangentially will experience infinite tidal forces in the radial direction when its speed relative to static observers is arbitrarily close to the speed of light. In \se{sec-spher}, we compute the tidal acceleration in a spherically symmetric spacetime that is geodesically complete but contains a central singularity. We show that the tidal acceleration is divergent at the central singularity, indicating that the curvature singularity may prevent the particle from reaching the origin. In addition, we find that the tidal acceleration of null geodesics similarly diverges at the origin. In \se{sec-wormhole}, we revisit a wormhole spacetime generated by a spherically symmetric electric field \cite{wormhole}. For $\delta_1\neq \delta_c$, this solution exhibits a curvature singularity at the throat of the wormhole. Although there is an obvious coordinate singularity at the throat, the authors claim that the spacetime is geodesically complete based on the analysis of geodesic behavior around the throat. We demonstrate explicitly that no well-defined metric exists at the throat, thereby proving geodesically incompleteness.
Furthermore,  we show that the tidal acceleration experienced by a particle approaching the singularity is always divergent. Concluding remarks are made in \se{sec-con}.

\section{Tidal force on the Earth's surface} \label{sec-earth}

Consider a static, spherically symmetric spacetime described by the metric
\bean
ds^2=-f(r)dt^2+\frac{1}{f(r)}dr^2+r^2d\theta^2+r^2\sin^2\theta d\phi^2 \,. \label{frds}
\eean
The static observer has the four-velocity
\bean
U^a=\frac{1}{\sqrt{f(r)}}\ppa{t}{a} \,.
\eean

Let $Z^a$ denote the unit tangent to a family of timelike geodesics and $W^a$ be a deviation vector field along the geodesics satisfying $Z^b\grad_b W^a=W^b\grad_b Z^a$. Here $W^a$ represents the displacement to an infinitesimally nearby geodesic in the family.
The relative acceleration of the nearby geodesic is then given by
\bean
A^a=Z^c\grad_c(Z^b\grad_b W^a)  \label{aaor}\,,
\eean
from which, we obtain the familiar expression \cite{waldgr}
\bean
A^a=R_{bcd}\hsp^a W^bZ^cZ^d  \label{aarw} \,,
\eean
where $R_{bcd}\hsp^a$ is the Riemann tensor. Note that, although $Z^a$ and $W^a$ are defined as vector fields in \eq{aaor}, they become local quantities in \eq{aarw} since \eq{aarw} obviously applies at any individual spacetime point. We see from \eq{aarw} that the relative acceleration is caused by spacetime curvature. Thus, $A^a$ is also called the tidal gravitational acceleration. For example,  given two nearby point particles in a curved spacetime connected by the deviation vector $W^a$, we can use \eq{aarw} to calculate their tidal acceleration or equivalently the tidal force, by multiplying the mass of one particle. \eq{aaor} makes it evident that the tidal acceleration can be computed point by point without knowledge of the particle's trajectory.

From \eq{aarw}, we observe that extreme tidal acceleration can arise not only from huge components of the curvature but also from huge components of $Z^a$. This happens when the relative speed with respect to the static observer is near the speed of light. In the following, we shall show explicitly that tangential motions could result in infinite tidal accelerations while radial motions could not.

Let $Z^a$ be the tangent to a fiducial geodesic. Since the spacetime is spherically symmetric, it is sufficient to consider a geodesic moving in the equatorial plane $\theta=\pi/2$. Then $Z^a$ takes the general form at a given point $p$:
\begin{align}
    &\left.Z^a\right|_p = \frac{\gamma}{\sqrt{f(r)}}\left.\ppa{t}{a}\right|_p+\sqrt{\gamma^2(1-u_{\phi}^2)-1}\sqrt{f(r)}\left.\ppa{r}{a}\right|_p+\frac{\gamma u_{\phi}}{r}\left.\ppa{\phi}{a}\right|_p\,.\label{zap}
\end{align}
Note that $\gamma\equiv-Z^aU_a$ and $Z^aZ_a=-1$. Thus, $\gamma$ is simply the familiar Lorentz factor, $\gamma=1/\sqrt{1-u^2}$, where $u<1$ is the 3-velocity measured by the static observer. The expression of $Z^r$ requires $\gamma^2(1-u_{\phi}^2)\geq1$, i.e., $u\geq u_{\phi}$.

We then construct an orthonormal tetrad along the geodesic
\begin{align}
    (e_0)^a \equiv& Z^a,\\
    (e_1)^a \equiv& \sqrt{\lb\frac{1}{u_{\phi}^2+1/\gamma^2}-1\rb\frac{1}{f(r)}}\ppa{t}{a}+\sqrt{\frac{f(r)}{u_{\phi}^2+1/\gamma^2}}\ppa{r}{a},\\
    (e_2)^a \equiv& \frac{1}{r}\ppa{\theta}{a},\\
    (e_3)^a \equiv& \frac{u_{\phi}\gamma}{\sqrt{f(r)\lb u_{\phi}^2+1/\gamma^2\rb}}\ppa{t}{a}+\gamma u_{\phi}\sqrt{\lb\frac{1}{u_{\phi}^2+1/\gamma^2}-1\rb f(r)}\ppa{r}{a}\nonumber\\
    &+\frac{\sqrt{\gamma^2u_{\phi}^2+1}}{r}\ppa{\phi}{a}.
\end{align}
The separation vector can be written as
\bean\label{Wonearth}
W^a=W^i(e_i)^a,\ \ \ i=1,2,3,
\eean
with the components satisfying the normalization condition
\bean
\lb W^1\rb^2+\lb W^2\rb^2+\lb W^3\rb^2=1\,. \label{wott}
\eean
Here and in what follows, we have omitted the subscript $p$ for $W^a$ or any other vector. However, one should keep in mind that the vectors are defined only at a spacetime point.
It is straightforward to verify that $Z^aW_a=0$ and $W^aW_a=1$. Note that $W^a$ is not necessarily normalized and its actual length should be the proper distance between the nearby geodesics. The choice above is adopted merely for convenience when our goal is to determine whether the tidal acceleration diverges. When the specific value of the acceleration is of interest, one simply multiplies the final result by the proper length of the separation.
Then straightforward calculation shows that the tidal acceleration $A^a$ has the form
\begin{align}
    A^a =& R_{bcd}\hsp^aW^bZ^cZ^d\nonumber\\
    =& \lsb\frac{\gamma^2u_{\phi}^2}{2r}f'(r)-\frac{f''(r)}{2}\lb\gamma^2u_{\phi}^2+1\rb\rsb W^1(e_1)^a+\left\{\frac{\gamma^2u_{\phi}^2}{r^2}\lsb f(r)-1\rsb-\frac{f'(r)}{2r}\lb\gamma^2u_{\phi}^2+1\rb\right\}W^2(e_2)^a\nonumber\\
    &-\frac{f'(r)}{2r}W^3(e_3)^a.
\end{align}

For the Schwarzschild solution $f(r)=1-2M_e/r$, we have
\begin{align}
    |A| &= \sqrt{A^aA_a} = \frac{M_e}{r^3}\sqrt{\lb2+3\gamma^2 u_{\phi}^2\rb^2\lb W^1\rb^2+\lb1+3\gamma^2 u_{\phi}^2\rb^2\lb W^2\rb^2+\lb W^3\rb^2} \,.\label{maga}
\end{align}
It is clear that when $u_{\phi}=0$, i.e., when the particle moves purely in the radial direction, the magnitude of $A^a$ becomes independent of the particle's speed and therefore remains finite. In contrast, as long as $Z^a$ has non-radial component, $|A|$ can diverge for particles moving arbitrarily close to the speed of light.

Furthermore, the Kretschmann scalar of Schwarzschild spacetime is
\begin{align}
    K \equiv R_{abcd}R^{abcd} = \frac{48M^2}{r^6}\,. \label{krfr}
\end{align}
Combing \eqs{maga} and \meq{krfr}, we obtain
\begin{align}
    \sqrt{\frac{K}{48}}\leq |A| \leq \sqrt{\frac{K}{48}}\lb2+\frac{3\gamma^2u_{\phi}^2}{r^2}\rb \,. \label{ulb}
\end{align}
The lower bound is attained when $W^a = (e_3)^a$, and the upper bound is attained  when $W^a = (e_1)^a$. That is, $|A|$ reaches its maximal value when the separation vector points in the radial direction. This implies that, on the surface of the earth, the maximum tidal force always occurs along the radial direction, independent of the particle's motion.

 Given a fixed $\gamma$, meaning a fixed energy measured by the static observer, one may ask how the particle must move in order to maximize $|A|$. \eq{ulb} indicates that, for fixed $\gamma$, the maximum of $|A|$ occurs when $u_\phi$ takes its maximum value. According to the discussion below \eq{zap},  the maximum value of $u_\phi$ is $u$, which corresponds to purely tangential motion (parallel to the surface of the earth). Therefore, the tidal acceleration is maximized when the particle moves tangentially while the separation vector is in the radial direction.

 Finally, note that the lower bound of $|A|$ is proportional to  $\sqrt K$. Hence, near the central singularity of a Schwarzschild black hole, the tidal acceleration diverges inevitably, regardless of  the particle's motion.

Now we employ Newtonian  gravity to demonstrate the reasonableness of \eq{maga}. In Newtonian gravity, the gravity acceleration on earth is given by
\bean
g=\frac{M}{r^2}\,,
\eean
where we have set the gravitational constant $G=1$.
Consider two point separated along the radial direction by a distance $\Delta r$ ($\ll r$). The difference of their acceleration is then
\bean
\Delta g=2\frac{M}{r^3}\Delta r \,. \label{ngr}
\eean
Since $\Delta g$ represents the relative acceleration between the two particles, it is identified as the tidal acceleration. When the two particles are aligned radially, we have $u_\phi=W^2=W^3=0$ and $W^1=1$
in \eq{maga}. Then after being multiplied by  $\Delta r$, \eq{maga} reduces precisely to \eq{ngr}.

We now quantitatively estimate the tidal acceleration experienced by a proton accelerated to ultrarelativistic velocities near the Earth's surface. Assume that $u=u_{\phi}$ and $W^a=(e_1)^a$. Then in the high Lorentz factor regime ($\gamma\gg 1$), the magnitude of the tidal acceleration in SI units scales as
\bean
|A| = G\frac{M_e}{r^3}\lb2+\frac{3u^2}{1-u^2}\rb = G\frac{M_e}{r^3}\lb3\gamma^2-1\rb \sim G\frac{M_e\gamma^2}{r^3} \label{agamma}\,.
\eean
This is the tidal acceleration per unit length.
The tidal force on a proton is then estimated by
\bean
F\sim m_p|A|r_p= G\frac{M_e\gamma^2}{r^3} m_p r_p \,,
\eean
where $m_p$ and $r_p$ are the mass and radius of proton, respectively. In the Large Hadron Collider (LHC), protons can reach $99.999999\%$ of the speed of light, corresponding to $\gamma\sim 7000$. Substituting the values $m_p\sim 10^{-27}\mathrm{kg}$, $r_p\sim 10^{-15}\mathrm{m}$, the mass of the earth $M\sim 10^{24}\mathrm{kg}$, the radius of the earth $r\sim 10^6\mathrm{m}$, we have
\bean
F\sim 10^{-40} \mathrm{N}  \,.
\eean
According to \cite{proton}, the local force acting on the struck quark in a proton is on the order of $3\ \mathrm{GeV}/\mathrm{fm}\sim 10^5 \mathrm{N}$. Although the force is amplified by a factor of $\gamma^2\sim 10^7$ due to the Lorentz boost effect, it remains negligible and is far from being sufficient to disrupt the proton. Particles with much higher energies have been found in cosmic rays. Protons in cosmic rays can reach energies up to $10^{20}\mathrm{eV}$ corresponding to a Lorentz factor of $\gamma\sim10^{11}$. Even in this case, the resulting tidal force is only $F\sim10^{-25}\mathrm{N}$, which remains far too small to be significant. We expect that tidal forces on elementary particles may have detectable effects in future experiments or observations.

\section{Tidal force at the center of a spherically symmetric spacetime} \label{sec-spher}

\subsection{Curvature singularity and tidal force}\label{sec2}

In the previous section, we have shown that ultrarelativistic particles can experience extremely high tidal forces even in the presence of a weak gravitational field. In this section, we shall investigate an alternative mechanism that gives rise to divergent tidal forces. This mechanism relies on curvature singularity, instead of the particle's speed.

The Computation of the Kretschmann curvature from metric \meq{frds} yields
\bean
K=R_{abcd}R^{abcd}=\frac{4-8 f+4f^2+4r^2f'^2+r^4f''^2}{r^4} \,.\label{krr}
\eean
We are interested in a regular metric at the coordinate origin $r=0$ that may be assumed in the form
\bean\label{expandf}
f(r)=1+C_1 r +C_2 r^2+C_3r^3+\mathcal{O}\lb r^4\rb \,.
\eean
Note that the leading term must be exactly unity  to avoid introducing a conical singularity at the origin ($r=0$).
Then \eq{krr} becomes
\bean
K=\frac{8 C_1^2}{r^2}+\frac{24C_1 C_2}{r}+\mathcal{O}\lb r^0\rb \,.
\eean
Thus, the necessary and sufficient condition for $K$ being divergent is $C_1\neq0$. In this case, the metric is regular at $r=0$ while the curvature is singular.
The stress-energy tensor $T_{ab}$ can be obtained via Einstein's equations and then the energy density measured by the static observers is given by
\bean
\rho=T_{ab}U^aU^b=-\frac{C_1}{4\pi r}-\frac{3C_2}{8\pi}+
\mathcal{O}\lb r\rb \,.
\eean
Thus, the energy density is also divergent as $K$ is divergent at the origin.

We shall check if a particle passing through the origin can suffer infinite tidal acceleration. Without loss of generality, we consider that the particle moves in the equatorial plane $\theta=\pi/2$ with the four-velocity
\bean
Z^a=\dot t \ppa{t}{a}+\dot r\ppa{r}{a}+\dot\phi\ppa{\phi}{a} \,,
\eean
where the ``dot'' represents the derivative with respect to the proper time of the particle. The conservation of energy and angular momentum yields
\bean
E\eqn-Z_a\ppa{t}{a} \label{eza} \,, \\
L\eqn Z_a\ppa{\phi}{a}  \,. \label{lza}
\eean
Along with the normalization condition $Z^aZ_a=-1$, we can solve for the components  of $Z^a$ as
\begin{align}
    &\dot t = \frac{E}{f(r)},\\
    &\dot \phi = \frac{L}{r^2},\\
    &\dot r^2 = E^2-V(r),\ \ V(r) \equiv f(r)\lb1+\frac{L^2}{r^2}\rb. \label{rde}
\end{align}
We choose the orthonormal tetrad to be
\begin{align}
    &(e_0)^a \equiv Z^a = \frac{E}{f(r)}\ppa{t}{a}-\sqrt{E^2-V(r)}\ppa{r}{a}+\frac{L}{r^2}\ppa{\phi}{a},\\
    &(e_1)^a \equiv\frac{E L}{f(r) \sqrt{L^2+r^2}}\ppa{t}{a}-L \sqrt{\frac{E^2-V(r)}{L^2+r^2}}\ppa{r}{a}+\frac{\sqrt{L^2+r^2}}{r^2}\ppa{\phi}{a},\\
    &(e_2)^a \equiv \frac{1}{r}\ppa{\theta}{a},\\
    &(e_3)^a \equiv -\frac{1}{f(r)}\sqrt{\frac{f(r)\lsb E^2-V(r)\rsb}{V(r)}}\ppa{t}{a}+\frac{Er}{\sqrt{r^2+L^2}}\ppa{r}{a}.
\end{align}
We take the deviation vector to be
\bean\label{Wfr}
W^a = W^i(e_i)^a,\ \ \ \sum_{i=1}^3\lb W^i\rb^2 = 1 \,,
\eean
which satisfies $W^aW_a=1$ and $W^aZ_a=0$. It is then straightforward to calculate the tidal acceleration as
\begin{align}
&A^a = \sum_{i=1}^3\Tilde{A}^iW^i(e_i)^a,\ \ |A| = \sqrt{\sum_{i=1}^3\lb \Tilde{A}^iW^i\rb^2},\ \ \  i=1,2,3  \,,\label{afr}\\
&\Tilde{A}^1 = -\frac{f'(r)}{2r},\\
&\Tilde{A}^2 = -\frac{f'(r)}{2r}-L^2\frac{rf'(r)+2\lsb1-f(r)\rsb}{2r^4},\\
&\Tilde{A}^3 = -\frac{f''(r)}{2}+L^2\frac{f'(r)-rf''(r)}{2r^3}.
\end{align}
It's clear that $|A|$ is finite for any $W^a$ in \eq{Wfr} is equivalent to that $\Tilde{A}^1, \Tilde{A}^2, \Tilde{A}^3$ are all finite.

We first consider the radial motion, i.e., $L=0$. For the regular metric given in \eq{expandf}, straightforward computation yields
\begin{align}\label{arad}
    \Tilde{A}^1 = \Tilde{A}^2 = -\frac{f'(r)}{2r} = -\frac{C_1}{2r}+\mathcal{O}\lb r^0\rb,\ \ \ \ \Tilde{A}^3 = -\frac{f''(r)}{2} = -C_2+\mathcal{O}\lb r\rb.
\end{align}

If the Kretschmann curvature $K$ is finite, i.e., $C_1=0$ as we have found above, we see that $|A|$ is finite at the origin.

If $C_1\neq 0$, it is easy to see that both $K$ and $|A|$ are divergent at the origin. Thus, for a particle moving in the radial direction, a divergent $K$ at the origin always implies a divergent $|A|$.

Now we turn to non-radial motions with $L\neq 0$. Recall \eq{rde}
\bean
\dot r^2=E^2-V(r),\ \ \ V(r)=f(r)\lb1+\frac{L^2}{r^2}\rb \label{vrb} \,.
\eean

Since $f(r)>0$, we see that the particle cannot reach the origin due to the infinite potential at $r=0$. However, if $L$ is chosen to be sufficiently small, the particle can be sufficiently close to the origin. Since we are interested in the region around the origin, it is enough to take the approximation $f\approx 1$. Denote $\epsilon$ the minimum distance to the origin that the particle can reach. Noting that $\dot r=0$ when $r=\epsilon$, we have
\bean
\epsilon=\sqrt{\frac{1}{E^2-1}}L \label{epel} \,.
\eean

For $C_1=0$, i.e., $K$ is finite, \eq{afr} becomes
\begin{align}
&\Tilde{A}^1|_{r=\ep} = -C_2+\mathcal{O}\lb\ep\rb\ ,\\
&\Tilde{A}^2|_{r=\ep} = \lsb-C_2+\mathcal{O}\lb\ep\rb\rsb+L^2\lsb-\frac{C_3}{2 \ep}+\mathcal{O}\lb\ep^0\rb\rsb \sim C_2,\nonumber\\
&\Tilde{A}^3|_{r=\ep} = \lsb-C_2+\mathcal{O}\lb\ep\rb\rsb+L^2\lsb-\frac{3C_3}{2 \ep}+\mathcal{O}\lb\ep^0\rb\rsb \sim -C_2.\nonumber
\end{align}
We see immediately that $|A|$ is finite.

For $C_1\neq 0$ ($K$ is divergent at $r=0$), we have
\begin{align}
    &\Tilde{A}^1|_{r=\ep} = -\frac{C_1}{2\ep}+\mathcal{O}\lb\ep^0\rb\, \label{acl},\\
    &\Tilde{A}^2|_{r=\ep} = \lsb-\frac{C_1}{2\ep}+\mathcal{O}\lb\ep^0\rb\rsb+L^2\lsb\frac{C_1}{2\ep^3}-\frac{C_3}{2 \ep}+\mathcal{O}\lb\ep^0\rb\rsb \sim \frac{C_1}{\ep},\nonumber\\
    &\Tilde{A}^3|_{r=\ep} = \lsb-C_2+\mathcal{O}\lb\ep\rb\rsb+L^2\lsb\frac{C_1}{2\ep^3}-\frac{3C_3}{2 \ep}+\mathcal{O}\lb\ep^0\rb\rsb \sim \frac{C_1}{\ep}.\nonumber
\end{align}

Since $\epsilon\sim L$ by \eq{epel}, $|A|$ is infinite at $r=0$ as long as $K$ is infinite.

In summary, for a static spherically symmetric spacetime with a regular metric given in \eq{expandf}, a finite Kretschmann scalar $K$ ensures that the tidal acceleration remains finite at the origin. Conversely, a divergent $K$ indicates that the tidal acceleration also is infinite.

\subsection{An example}
The regular black hole soltuion with a nonlinear electrodynamics source, as analyzed in \cite{regume}, is described by the metric function
\bean
f(r)=1-\frac{2m r}{q^2+r^2} \,,
\eean
where $m$ and $q$ represent the black hole's mass and electric charge, respectively. This metric function ensures both regularity at $r=0$ and asymptotic flatness at infinity. Through Taylor expansion about $r=0$, we have
\bean
f(r\sim 0)=1-\frac{2mr}{q^2}+\frac{2m r^3}{q^4}+\mathcal{O}(r^5) \,.
\eean
Obviously, this metric is regular at $r=0$ with $C_1=-\frac{2m}{q^2}$. The Kretschmann curvature is
\bean
K=\frac{16 m^2 \left(16 m^4 r^4-32 m^3 q^2 r^3+29 m^2 q^4 r^2-12 m q^6 r+2 q^8\right)}{r^2 \left(q^2-2 m r\right)^6} \,,
\eean
which is divergent at $r=0$ as expected.

We first consider the radial motion ($L=0$). \eq{rde} becomes
\bean
\dot r^2=E^2-\left( 1-\frac{2m r}{q^2+r^2}\right) \,.
\eean
Thus, the particle can reach and pass through the origin. According to \eq{arad}, the tidal acceleration diverges at the origin as
\bean
\Tilde{A}^1 = \Tilde{A}^2 \sim -\frac{m}{q^2r} \,.
\eean

\begin{figure}[htbp]
\centering
\includegraphics[width=9cm]{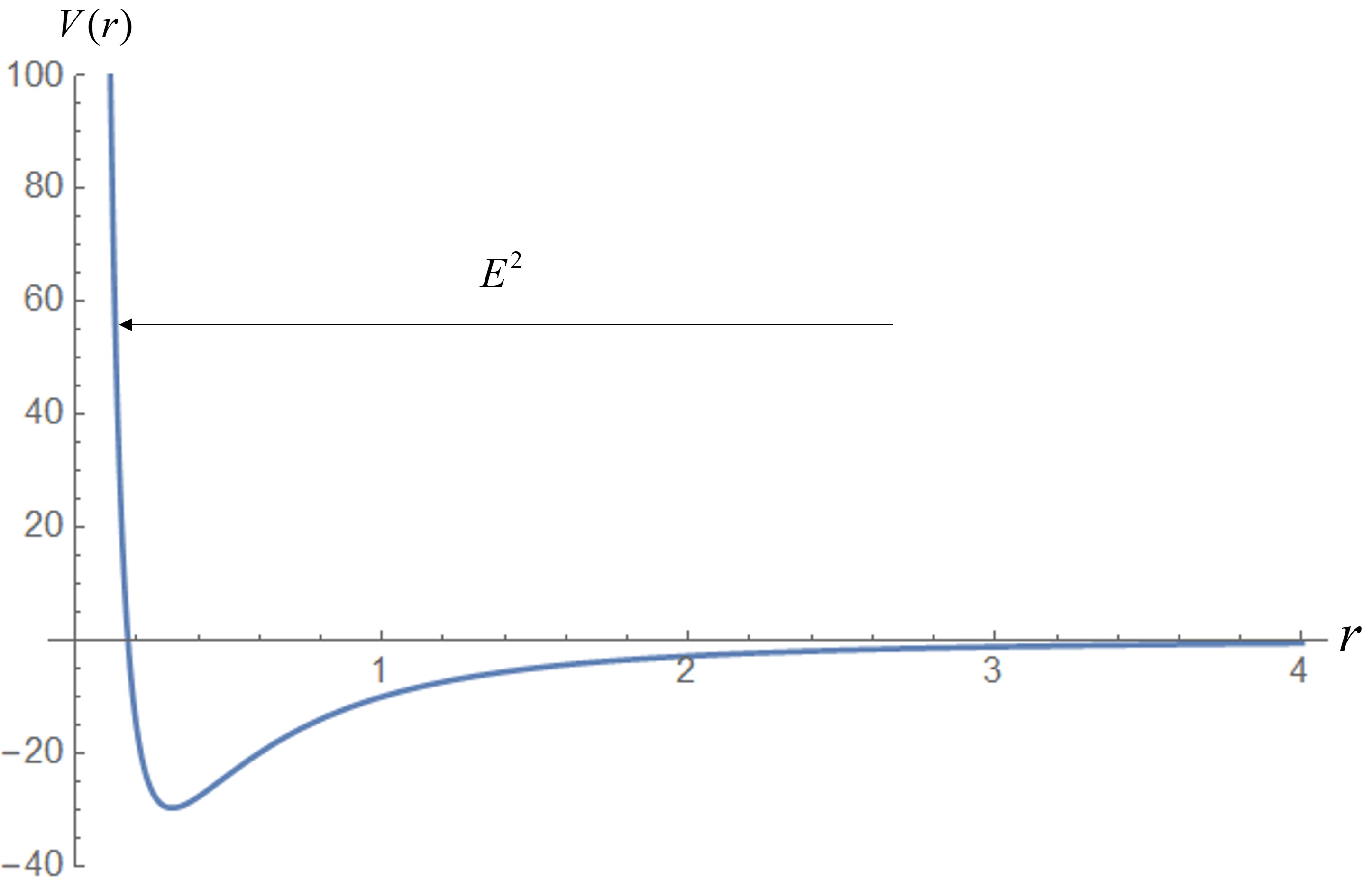} 
\caption{For nonradial motion, the potential $V(r)$ goes to infinity at $r=0$.}\label{fig-vrr}
\end{figure}

For non-radial motion, the effective potential in \eq{vrb} is described by \fig{fig-vrr}. It shows that an infinite potential barrier always exists at $r=0$ and then no particle can reach it. According to \eqs{epel} and \meq{acl}, $A^1,A^2,A^3$ diverges as $ 1/\epsilon$.

\subsection{Null geodesics}

Now we examine geodesic deviation of null geodesics. Let $k^a$ be the tangent of a null geodesic in the equatorial plane. By replacing $Z^a$ with $k^a$ in \eqs{eza} and \meq{lza}, we can define the conserved energy $E$ and angular momentum $L$ for the null geodesic. Together with the null condition $k^ak_a=0$, we have
\begin{align}
    k^a = \frac{E}{f(r)}\ppa{t}{a}-\sqrt{E^2-f(r)\frac{L^2}{r^2}}
    \ppa{r}{a}+\frac{L}{r^2}\ppa{\phi}{a} \,.
\end{align}
Analogous to the timelike case, the tidal acceleration of null geodesics is defined by
\bean
A^a=k^c\grad_c(k^b\grad_b W^a) \,, \label{kor}
\eean
where  $W^a$ is the deviation vector satisfying ${\cal L}_k W^a=0$ and $k^aW_a=0$.  It follows from \eq{kor} that
\bean
A^a=R_{bcd}\hsp^a W^bk^ck^d  \label{krw} \,.
\eean
It is evident from \eq{kor} that $A^a$ represents the relative acceleration of two nearby null geodesics. Unlike timelike geodesics, null geodesics exhibit purely relative acceleration without an associated ``tidal force'' interpretation.

For a radial null geodesic($L=0$), we see
\begin{align}
    A^a = R_{bcd}\hsp^a W^bk^ck^d = \frac{f''(r)}{2}W^bk_bk^a = 0.
\end{align}

For $L\neq0$, define two spacelike vector fields
\bean
   (e_1)^a \eqn \frac{1}{r}\ppa{\theta}{a},\\
    (e_2)^a \eqn -\frac{r}{Lf(r)}\sqrt{E^2-f(r)\frac{L^2}{r^2}}\ppa{t}{a}+\frac{E r}{L}\ppa{r}{a} \,,
\eean
which satisfy
\bean
    k^a(e_i)_a = 0,\qquad(e_i)^a(e_j)_a = \delta_{ij}\,, \qquad i,j=1,2 \,.
\eean
Thus, $\{(e_1)^a,(e_2)^a\}$ form a two dimensional space orthogonal to $k^a$. A general deviation vector can be expressed as
\bean
W^a=\alpha (e_1)^a+\beta (e_2)^a \,.
\eean

For the regular metric in the form of \eq{expandf}, we have
\begin{align}
    A^a &= \Tilde{A}^1\alpha(e_1)^a+\Tilde{A}^2\beta(e_2)^a\nonumber\\
    &=\frac{L^2}{2}\left\{-\frac{rf'(r)-2f(r)+2}{r^4}\alpha(e_1)^a
    +\frac{f'(r)-rf''(r)}{r^3}\beta(e_2)^a\right\}\,,\\
    \Tilde{A}^1|_{r=\ep} &= \frac{L^2}{2}\lsb\frac{C_1}{\ep^3}-\frac{C_3}{\ep}+\mathcal{O}\lb r^0\rb\rsb,\\
    \Tilde{A}^2|_{r=\ep} &= \frac{L^2}{2}\lsb-\frac{C_1}{\ep^3}+\frac{3C_3}{\ep}+\mathcal{O}\lb r^0\rb\rsb.
\end{align}
The radial equation
\bean
\dot r^2=E^2-f(r)\frac{L^2}{r^2}
\eean
gives the minimum distance to the origin
\bean
\epsilon\sim L \,.
\eean
Therefore, $\Tilde{A}^1,\ \Tilde{A}^2$ diverges near the origin as $1/\epsilon$.

\section{Tidal forces near a wormhole} \label{sec-wormhole}

In contrast to conventional spherically symmetric spacetimes with a central origin at $r=0$, wormhole geometries are characterized by a minimum surface called the throat. In this section, we explore the nature of singularity at the throat.

\subsection{The wormhole singularity}
Olmo, et at. investigated a spherically symmetric wormhole generated by the interplay between the electric field and the Palatini gravity \cite{wormhole}.
The corresponding metric has the form
\begin{align}
    \md s^2 = -A(x)\md t^2+\frac{1}{A(x)\sigma_+^2}\md x^2+r^2(x)\md\Omega^2, \label{dsax}
\end{align}
where
\begin{align}
    &A(x) = \frac{1}{\sigma_+}\left[1-\frac{r_S}{r}\frac{1+\delta_1G(r/r_c)}{\sigma_-^{1/2}}\right],\\
    &\delta_1 = \frac{1}{2r_S}\sqrt{\frac{r_q^3}{l_{\ep}}},\ \ \ \ \ r_c = \sqrt{l_{\ep}r_q},\\
    &\sigma_{\pm} = 1\pm\frac{r_c^4}{r^4(x)}, \label{sigrc} \\
    &r^2(x) = \frac{x^2+\sqrt{x^4+4r_c^4}}{2}\,. \label{rtx}
\end{align}
Here $r_S=2M_0$ is Schwarzschild radius, and  the function $G(z)$ takes the form
\begin{align}
    &G(z) = -\frac{1}{\delta_c}+\frac{\sqrt{z^4-1}}{2}\lsb f_{\frac{3}{4}}(z)+f_{\frac{7}{4}}(z)\rsb,\\
    &f_{\lambda}(z) = {}_2F_1\lb\frac{1}{2},\ \lambda,\ \frac{3}{2},\ 1-z^4\rb,
\end{align}
where $\delta_c>0$ is a constant.

The throat of the wormhole is located at $r=r_c=\sqrt{l_\epsilon r_q}$, where $l_\epsilon$ and $r_q$ characterize the high-curvature corrections and the length scale associated with the electric charge, respectively. Note that $x=0$ at $r=r_c$.

The Krestschmann scalar is computed to be
\bean
 K &=&
  \frac{1}{4 r^4}\left\{8 A^2 \left[r^2 \left(\sigma _-'\right){}^2+2 \sigma _-^2\right]+r^4 \left(A'\right)^2 \left(\sigma _-'\right){}^2+16 A \sigma _- \left(r^2 A' \sigma _-'-2\right)+4 r^4 \sigma _- A' A'' \sigma _-'\right.
    \non
   & & \left. +4 \sigma _-^2 \left[r^4 \left(A''\right)^2+4 r^2 \left(A'\right)^2\right]+16\right\}  \,, \label{kworm}
\eean
where the prime denotes the radial derivative $\pp{}{r}$.
Near the throat, $A(x)$ exhibits the asymptotic hehavior \cite{wormhole}
\begin{align}
    \lim_{r\to r_c}A(x) = \frac{N_q}{4N_c}\frac{\delta_1-\delta_c}{\delta_1\delta_c}
    \sqrt{\frac{r_c}{r-r_c}}+\frac{N_c-N_q}{2N_c}
    +\mathcal{O}\lb\sqrt{r-r_c}\rb, \label{asya}
\end{align}
where $N_q$ is the number of charges defined by $N_q=q/e$ and $N_c$ is a constant. Substitution of \eq{asya} into \eq{kworm} yields
 \begin{align}
    K   =& \frac{\left(\delta _c-\delta _1\right)^2 r_S^2}{4 \delta _c^2 r_c^3 \left(r-r_c\right)^3}-\frac{5 \left(\delta _c-\delta _1\right)^2 r_S^2}{8\delta _c^2 r_c^4 \left(r-r_c\right)^2 }+\frac{2 \left(\delta _c-\delta _1\right) r_S \left(2 \delta _1 r_S-3 r_c\right)}{3 \delta _c r_c^{9/2} \left(r-r_c\right)^{3/2}}-\frac{11 \left(\delta _c-\delta _1\right)^2 r_S^2}{8\delta _c^2 r_c^5 \left(r-r_c\right)}\non
    &+\frac{\left(\delta _c-\delta _1\right) r_S \left(435 r_c-418 \delta _1 r_S\right)}{30 \delta _c r_c^{11/2} \sqrt{r-r_c}}+\mathcal{O}\lsb\lb r-r_c\rb^0\rsb.
\end{align}

For $\delta_1=\delta_c$, the metric function $A(x)$ is regular at $r=r_c$ and $K$ is finite. Hence, the metric is well-defined and free from curvature singularities at the throat. However, this particular case is not the focus of our investigation.

For $\delta_1\neq \delta_c$, both $A(x)$ and $K$ are divergent at the throat. By investigating the behavior of radial geodesics near the throat, it is claimed that the spacetime is geodesically complete and  particles can traverse the throat \cite{wormhole}\cite{Olmo:2016fuc}\cite{Olmo:2015dba}. We find that this is not true because that the metric cannot be defined at the throat. The singular behavior of $A(x)$ at the throat does not necessarily mean that $r=r_c$ is a spacetime singularity. It may be a coordinate singularity like that on the horizon of Schwarzschild black hole. If it were a coordinate singularity, we should be able to find a coordinate transformation such that the new coordinates can cover $r=r_c$ and the metric components are all regular. We shall take the case where $\delta_1>\delta_c$ as an example to show that no such transformation exists. For this purpose, we first expand \eq{rtx} around the throat ($x=0$) and find
\bean
r=r_c+\frac{x^2}{4r_c}+\mathcal{O}(x^4)\,.
\eean
Then near the throat, \eq{asya} becomes
\bean
A(x)\approx k x^{-1}\,,
\eean
where
\begin{align}
    k = \frac{N_q}{2N_c}\frac{\delta_1-\delta_c}{\delta_1\delta_c}r_c
\end{align}
Consequently, \eq{dsax} can be written as
\begin{align}
    \md s^2 = -kx^{-1}\md t^2+\frac{x}{k \sigma_+^2}\md x^2 +r^2d\Omega^2\,.  \label{dskx}
\end{align}
By virtue of \eq{sigrc}, we see that $\sigma_+\approx 2$ near the throat and thus can be treated as a constant.

Rewrite \eq{dskx} as
\begin{align}
    \md s^2 = kx^{-1}\left( -\md t^2+\frac{x^2}{k^2 \sigma_+^2}\md x^2 \right) \,.\label{ndsk}
\end{align}
Here, we have omitted the angular components $(\theta,\phi)$ of the metric as they remain unaffected by the coordinate transformation. This reduction makes the relevant spacetime is effectively two-dimensional.

Define the null coordinates
\bean
u\eqn t-\frac{1}{2k\sigma_+}x^{2} \,,\\
v\eqn t+\frac{1}{2k\sigma_+}x^{2}\,.
\eean
In these coordinates, the metric takes the form
\bean
\md s^2=-\alpha \left(\frac{v-u}{2}\right)^{-1/2}\md u \md v \,,\label{vmu}
\eean
where $\alpha=\sqrt{k/(2\sigma_+)}$, which can be treated as a nonzero constant near the throat. Hence, without loss of generality, we shall take $\alpha=1$ in the following calculation.
Obviously, the coordinate singularity $x=0$ corresponds to $u=v$. As evident from \eq{vmu}, the metric remains singular in these coordinates.
If the metric is regular at the throat, we can always find another set of null coordinates $(U,V)$ such that
\bean
\md s^2=-F(U,V)\md U \md V \,,
\eean
where $F$ is a regular, nonvanishing function. The two coordinate systems are related by the transformation
\bean
U\eqn U(u,v)\,,  \\
V\eqn V(u,v)\,.
\eean
For the two-dimensional spacetime, there are only two null directions. Consequently, we have
\bean
\ppa{U}{a} // \ppa{u}{a}, \ \ \ \  \ppa{V}{a}// \ppa{v}{a} \,,
\eean
which implies $\partial_vU=\partial_uV=0$. Hence, the metric transforms as
\bean
\md s^2=-F(U,V)U'(u)V'(v) \md u \md v \,.
\eean
Comparting with \eq{vmu}, we obtain
\bean
F(U,V)U'(u)V'(v)=(v-u)^{-1/2} \,. \label{fuvu}
\eean
This can be equivalently expressed as
\bean
F(U,V)^{-2}[U'(u)V'(v)]^{-2}=v-u \,.\label{bfuv}
\eean
Since the right-hand side of \eq{bfuv} vanishes at the throat $u=v$, we have
\bean
U'(u)V'(u)\to \infty
\eean
for $-\infty<u<\infty$. Obviously, no functions satisfy this divergent behavior everywhere.

Therefore, one cannot eliminate the coordinate singularity at $r=r_c$ by any coordinate transformation, meaning that no metric can be defined at the throat. Consequently, $r=r_c$ is a true spacetime singularity. Geodesics reach the singularity in finite proper time. Thus, the spacetime is geodesically incomplete. For $\delta_1\neq \delta_c$, the wormhole structure fails to exist as the spacetime terminates at $r=r_c$.

\subsection{Tidal acceleration near the singularity}

Now the spacetime is confined in the region $x>0$. Since the curvature diverges as $x\to 0$, it is of interest to check whether the tidal acceleration is divergent as a particle approaches $x=0$.

The equations of motion of a timelike geodesic in the equatorial plane are given by
\begin{align}
    &\dot t = \frac{E}{A},\\
    &\dot \phi = \frac{L}{r^2},\\
    &\dot x^2 = \sigma_+^2\lsb E^2-A\lb1+\frac{L^2}{r^2}\rb\rsb = \sigma_+^2\lsb E^2-V(r)\rsb\,, \label{xdots}
\end{align}
where
\begin{align}
    V (r)= A\lb1+\frac{L^2}{r^2}\rb = \frac{\left(\delta _1-\delta _c\right) r_S \left(r_c^2+L^2\right)}{4 \delta _c r_c^{5/2} \sqrt{r-r_c}}+\mathcal{O}\lsb\lb r-r_c\rb^0\rsb.  \label{vral}
\end{align}

For $\delta_1<\delta_c$, $V(r)<0$ and then there is an infinitely  attractive potential at $x=0$. Consequently, particles can easily reach the throat.

For $\delta_1>\delta_c$, the potential becomes infinitely repulsive at the throat, preventing particles from reaching it. Instead, an incoming particle is deflected and orbits the wormhole at a minimum radius $r_m (>r_c)$. Since $\dot x$ vanishes at the turning point $r=r_m$, we can combine \eqs{xdots} and \meq{vral} to obtain
\begin{align}
E^2=V(r_m)=A|_{r=r_m}\lb1+\frac{L^2}{r_m^2}\rb\sim \frac{1}{\sqrt{r_m-r_c}}+\mathcal{O}\lsb\lb r_m-r_c\rb^0\rsb \,,
\end{align}
which shows that the $r_m-r_c\sim E^{-4}$. Since the energy $E$ is finite, the particle cannot get arbitrarily close to the throat.

Our goal is to determine whether the tidal acceleration is infinite near the throat. The four-velocity of a timelike particle is given by
\begin{align}\label{geo}
    Z^a = \frac{E}{A}\ppa{t}{a}-\sigma_+\sqrt{ E^2-V}\ppa{x}{a}+\frac{L}{r^2}\ppa{\phi}{a}\,.
\end{align}

We choose the separation vector $W^a$ as
\begin{align}\label{Wworm}
    W^a = \frac{1}{r}\ppa{\theta}{a}.
\end{align}
We first analyze the case $\delta_1<\delta_c$, where, as previously discussed, the particle can reach the throat when $\delta_1<\delta_c$. The tidal acceleration near the throat is given by
\begin{align}
    &A^a = \Tilde{A}\ W^a \,\\
    &\Tilde{A} = \frac{\left(\delta _c-\delta _1\right) r_S \left(r_c^2+L^2\right)}{4 \delta _c r_c^{9/2} \sqrt{r-r_c}}+\mathcal{O}\lsb\lb r-r_c\rb^0\rsb\,.\label{a2}
\end{align}
Clearly, the tidal acceleration associated with separation direction \eq{Wworm} diverges as $r\to r_c$. In the case $\delta_1>\delta_c$, we have shown that the particle with finite energy cannot approach arbitrarily close to the throat. Consequently, its tidal acceleration remains finite for any deviation vector.

An interesting result obtained in \cite{Olmo:2016fuc}\cite{Olmo:2015dba} is that, despite the divergence of the tidal acceleration, the proper distance between two adjacent geodesics is always finite near the singularity. The authors also show that a scalar field propagating in this background is well-behaved everywhere. These results indicate that the curvature singularity may not be so destructive for extended bodies approaching it.

\section{Concluding remarks} \label{sec-con}
In this paper, we have explored the relations among geodesic completeness, curvature singularities and tidal forces. Our key findings are as follows:

We demonstrated that the tidal force experienced by a particle could be extremely large even on the Earth. To achieve this, the particle must move non-radially at ultrarelativistic speed. This result illustrates that ultrarelativistic particles can amplify the magnitude of gravitational field. We found that the maximum tidal force experienced by a particle is always along the radial direction, regardless of the particle's velocity. Moreover, at a fixed particle energy, purely tangential motion produces the largest tidal acceleration. For the Schwarzschilds solution, a lower bound on the tidal acceleration has been found, which is proportional to the square root of the the Krestschmann curvature. This result shows that while the Lorentz boot effect can make the tidal acceleration arbitrarily large, it cannot make it arbitrarily small in regimes of strong gravity.

For a class of spherically symmetric spacetimes with metric regular at the origin but harboring curvature singularities, we showed that a particle moving in the radial direction can pass through the origin and endure infinite tidal forces. For non-radial motions, by fine-tuning the  angular momentum to arbitrarily small values, particles can approach arbitrarily close to the origin, where the tidal forces likewise diverge. If a photon travels strictly in the radial direction, its tidal acceleration associated with any deviation vector vanishes at the origin. Otherwise, the tidal acceleration for a photon diverges near the origin.

Finally, we reexamined a wormhole solution that was claimed to be geodesically complete and possesses a curvature singularity at its throat. Our analysis reveals that when $\delta_1\neq \delta_c$, the coordinate singularity at the throat cannot be eliminated by coordinate transformation and thus the solution is geodesically incomplete. This incompleteness implies that the spacetime terminates at $r=r_c$, preventing the formation of a traversable wormhole geometry. We also deomonstrated that particles always experience infinitely large tidal forces as they approach the singularity.

We have shown that tidal forces can be significantly enhanced by the factor $\gamma^2$ for ultarrelativistic particles. While this effect is currently undetectable due to energy limitations of particle colliders, ultrahigh-energy cosmic rays could make such tidal effects potentially observable in future experiments. Moreover, gravitational fields near ultra compact objects, such as neutron stars and black holes, are much stronger than that near the Earth, implying observable effects related to the amplified tidal forces.

Our analysis demonstrates that curvature singularities are generically accompanied by  divergent tidal accelerations even in geodesically complete spacetimes. We have specifically examined this relationship through the Kretschmann scalar in spherically symmetric geometries. It would be worthwhile to extend the discussion to other curvature invariants in more general spacetime backgrounds.

\section*{Acknowledgement}
We thank Prof. Robert Wald for valuable comments on the relationship between the tidal force and Lorentz boost as well as recommending the references \cite{mat74}-\cite{sexl71}. We also thank Prof. Xiaoning Wu for helpful discussions. This research was supported in part by NSFC Grants No. 12205013.

\end{document}